\documentclass[twoside,11pt]{article}

\usepackage[preprint]{jmlr2e}

\usepackage[utf8]{inputenc} 
\usepackage[T1]{fontenc}    
\usepackage{microtype}      

\usepackage{xcolor}

\usepackage{enumitem}
\setitemize{itemsep=2pt,topsep=2pt,parsep=2pt,partopsep=2pt} 
\setenumerate{itemsep=2pt,topsep=2pt,parsep=2pt,partopsep=2pt}

\usepackage{amsmath}
\usepackage{amssymb}
\usepackage{amsfonts}       
\usepackage{nicefrac}       
\usepackage{mathtools}
\usepackage{relsize}
\usepackage{bm}

\usepackage{etoolbox}
\usepackage{pgfplots}
\usepgfplotslibrary{groupplots}
\usepackage{tikz}
\usetikzlibrary{tikzmark}
\usetikzlibrary{calc}
\usetikzlibrary{fadings}
\usetikzlibrary{patterns}
\usetikzlibrary{shadows.blur}
\usetikzlibrary{shapes}

\usetikzlibrary{external}
\tikzset{external/force remake=false}
\tikzsetexternalprefix{figures/external/}

\newcommand{\colordot}[2][gray,fill=gray]{\tikz[baseline=-0.5ex]\draw[#1,radius=#2] (0,0) circle ;}%

\usepackage{booktabs}
\usepackage{siunitx}
\usepackage{multirow}

\usepackage{graphicx}
\usepackage{pgfplots}
\pgfplotsset{compat=1.15}
\usepackage{caption}
\usepackage[labelformat=simple]{subcaption}
\usepackage{adjustbox}

\usepackage{algorithm}
\usepackage[noend]{algpseudocode}
\algrenewcommand{\algorithmiccomment}[1]{\hfill {\small \textcolor{darkgray}{$\mathsmaller \vartriangleright$ #1}}}  
\algrenewcommand\algorithmicindent{2em}   
\algrenewcommand\alglinenumber[1]{\small {\textcolor{darkgray}{#1}}} 

\makeatletter
\expandafter\patchcmd\csname\string\algorithmic\endcsname{\itemsep\z@}{\itemsep=0.25ex}{}{}
\newcommand\fs@booktabsruled{%
  \def\@fs@cfont{\bfseries\strut}\let\@fs@capt\floatc@ruled
  \def\@fs@pre{\hrule height\heavyrulewidth depth0pt \kern\belowrulesep}%
  \def\@fs@mid{\kern\aboverulesep\hrule height\lightrulewidth\kern\belowrulesep}%
  \def\@fs@post{\kern\aboverulesep\hrule height\heavyrulewidth\relax}%
  \let\@fs@iftopcapt\iftrue
}
\makeatother
\floatstyle{booktabsruled}
\restylefloat{algorithm}
\usepackage{listings}
\definecolor{codegreen}{rgb}{0,0.6,0}
\definecolor{codegray}{rgb}{0.5,0.5,0.5}
\definecolor{codepurple}{rgb}{0.58,0,0.82}
\definecolor{codebackground}{rgb}{0.973,0.973,0.973}

\lstdefinestyle{myPythonStyle}{
    backgroundcolor=\color{codebackground},   
    basicstyle=\ttfamily\footnotesize,        
    breakatwhitespace=false,         
    breaklines=true,                 
    captionpos=b,                    
    commentstyle=\color{codegreen},    
    deletekeywords={...},            
    escapeinside={\%*}{*)},          
    extendedchars=true,              
    firstnumber=1000,                
    frame=single,                    
    keepspaces=true,                 
    keywordstyle=\color{blue},       
    language=Python,                 
    morekeywords={*,...},            
    numbers=none,                    
    numbersep=5pt,                   
    numberstyle=\tiny\color{codegray}, 
    rulecolor=\color{black},         
    showspaces=false,                
    showstringspaces=false,          
    showtabs=false,                  
    stepnumber=2,                    
    stringstyle=\color{codepurple},     
    tabsize=2,                       
}
\usepackage{tocloft}        
\usepackage[prependcaption,textsize=small,color=gray!40]{todonotes} 

\usepackage{url}
\usepackage{xr}
\usepackage{hyperref}
\hypersetup{
    colorlinks,
    linkcolor={black},
    citecolor={blue!50!black},
    urlcolor={blue!50!black}
}
\usepackage{cleveref}
\usepackage[numbered]{bookmark} 
\setcitestyle{sort&compress,numbers,square,comma}

\usepackage{amsmath}
\usepackage{amsfonts}
\usepackage{amssymb}
\usepackage{bm}
\usepackage{physics}

\renewcommand{\top}{{\intercal}}

\DeclareSymbolFont{stmry}{U}{stmry}{m}{n}
\DeclareMathSymbol\obar\mathrel{stmry}{"3A}
\DeclareMathSymbol\otimes\mathrel{stmry}{"0F}
\DeclareMathSymbol\ominus\mathrel{stmry}{"17}
\makeatletter
\newcommand{\superimpose}[2]{
  {\ooalign{$#1\@firstoftwo#2$\cr\hfil$#1\@secondoftwo#2$\hfil\cr}}}
\makeatother

\makeatother

\DeclareMathAlphabet{\mathsfit}{\encodingdefault}{\sfdefault}{m}{sl}
\SetMathAlphabet{\mathsfit}{bold}{\encodingdefault}{\sfdefault}{bx}{n}

\newcommand{\probnum}{\texttt{ProbNum}}
\newcommand{\papertitle}{\probnum: Probabilistic Numerics in Python}
\newcommand{\authors}{Jonathan Wenger, Nicholas Krämer, Marvin Pförtner, Jonathan Schmidt, Nathanael Bosch, Nina Effenberger, Johannes Zenn, Alexandra Gessner, Toni Karvonen, Fran\c{c}ois-Xavier Briol, Maren Mahsereci, Philipp Hennig}

\newif\ifarxiv
\arxivtrue

\ifarxiv
  \jmlrheading{}{\the\year{}}{}{}{}{\authors}
\else
\jmlrheading{1}{\the\year{}}{1-48}{7/21}{03/22}{wenger22}{\authors}
\fi

\ShortHeadings{{\papertitle}}{Wenger et al.}
\firstpageno{1}

\begin{document}

\title{\papertitle}

\author{\name Jonathan Wenger \email jonathan.wenger@uni-tuebingen.de\\
  \name Nicholas Kr\" amer \email nicholas.kraemer@uni-tuebingen.de\\
  \name Marvin Pförtner \email marvin.pfoertner@uni-tuebingen.de\\
  \name Jonathan Schmidt \email jonathan.schmidt@uni-tuebingen.de\\
  \name Nathanael Bosch \email nathanael.bosch@uni-tuebingen.de\\
  \name Nina Effenberger \email nina.effenberger@uni-tuebingen.de\\
  \name Johannes Zenn \email johannes.zenn@student.uni-tuebingen.de\\
  \name Alexandra Gessner \email alexandra.gessner@uni-tuebingen.de\\
  \addr University of Tübingen, Tübingen, Germany\\
  \name Toni Karvonen \email toni.karvonen@helsinki.fi\\
  \addr University of Helsinki, Helsinki, Finland\\
  \name Fran\c{c}ois-Xavier Briol \email f.briol@ucl.ac.uk\\
  \addr University College London, London, UK\\
  \name Maren Mahsereci \email maren.mahsereci@uni-tuebingen.de\\
  \name Philipp Hennig \email philipp.hennig@uni-tuebingen.de\\
  \addr University of Tübingen, Tübingen, Germany
}

\ifarxiv
  \editor{} 
\else
  \editor{TBA}
\fi

\maketitle

\ifarxiv
  \thispagestyle{plain} 
\fi

\begin{abstract}
  Probabilistic numerical methods (PNMs) solve numerical problems via probabilistic inference.
  They have been developed for linear algebra, optimization, integration and differential equation
  simulation. PNMs naturally incorporate prior information about a problem and quantify uncertainty
  due to finite computational resources as well as stochastic input. In this paper, we present
  \probnum{}: a Python library providing state-of-the-art probabilistic numerical
  solvers. \probnum\ enables custom composition of PNMs for specific problem classes via a modular
  design as well as wrappers for off-the-shelf use. Tutorials, documentation, developer guides and
  benchmarks are available online at
  \url{www.probnum.org}.
\end{abstract}

\begin{keywords}
  probabilistic numerics, machine learning, numerical analysis
\end{keywords}

\section{Introduction}

Mathematical models are used to explain and predict the behavior of complex systems in all fields
of engineering and the physical sciences. In practice, their application relies heavily on
numerical approximation. However, a finite
computational budget or noise-corrupted inputs induce uncertainty over the quality of any computed
approximation.
For example, the accurate simulation of a tsunami model may take several
hours \citep{Behrens2015}, but operational needs require much faster approximate simulation
\citep{Giles2020}. Quantification of the induced error is crucial to any practitioner
relying on the output of such a model.

\emph{Probabilistic numerics (PN)} \citep{Hennig2015a,Cockayne2019a,Oates2019} quantifies uncertainty in computation by
solving problems from numerical analysis with statistical inference.
The probabilistic viewpoint provides a principled way to encode structural knowledge about a
problem into the algorithm and gives rise to new functionality beyond the scope of
non-probabilistic methods \citep{Schmidt2021}.

Previously, implementations of PN algorithms have been either standalone, only available for a
specific numerical problem, or not available at all. This lack of quality open-source
software has inhibited the development of PN methods and their application. To promote the
widespread adoption of PN in the scientific community, we introduce \probnum{}, an
open-source Python library implementing probabilistic numerical methods. Our library
\begin{itemize}
  \item facilitates \emph{rapid experimentation} with, and \emph{application} of, state-of-the-art PN
        solvers;
  \item enables \emph{custom composition of solvers} for specific problems from predefined components;
  \item simplifies and accelerates \emph{development of new methods} via a unified API.
\end{itemize}
Finally, by implementing PN methods in a composable manner, \probnum{} is an initial step towards the vision of PN of propagating uncertainty through chains of computations \cite{Hennig2015a}.

\begin{figure}
  \begin{subfigure}[b]{0.54\textwidth}
    \centering
\begin{lstlisting}
import probnum as pn

belief = pn.linalg.problinsolve(A, b)
belief.x
# <Normal with shape=(n,), dtype=float64>
\end{lstlisting}
    \subcaption{Solving a linear system with \probnum{}.\label{fig:interface-example}}
  \end{subfigure}\hfill
  \begin{subfigure}[b]{0.44\textwidth}
    \centering
    \raisebox{0em}{\includegraphics[width=0.8\textwidth]{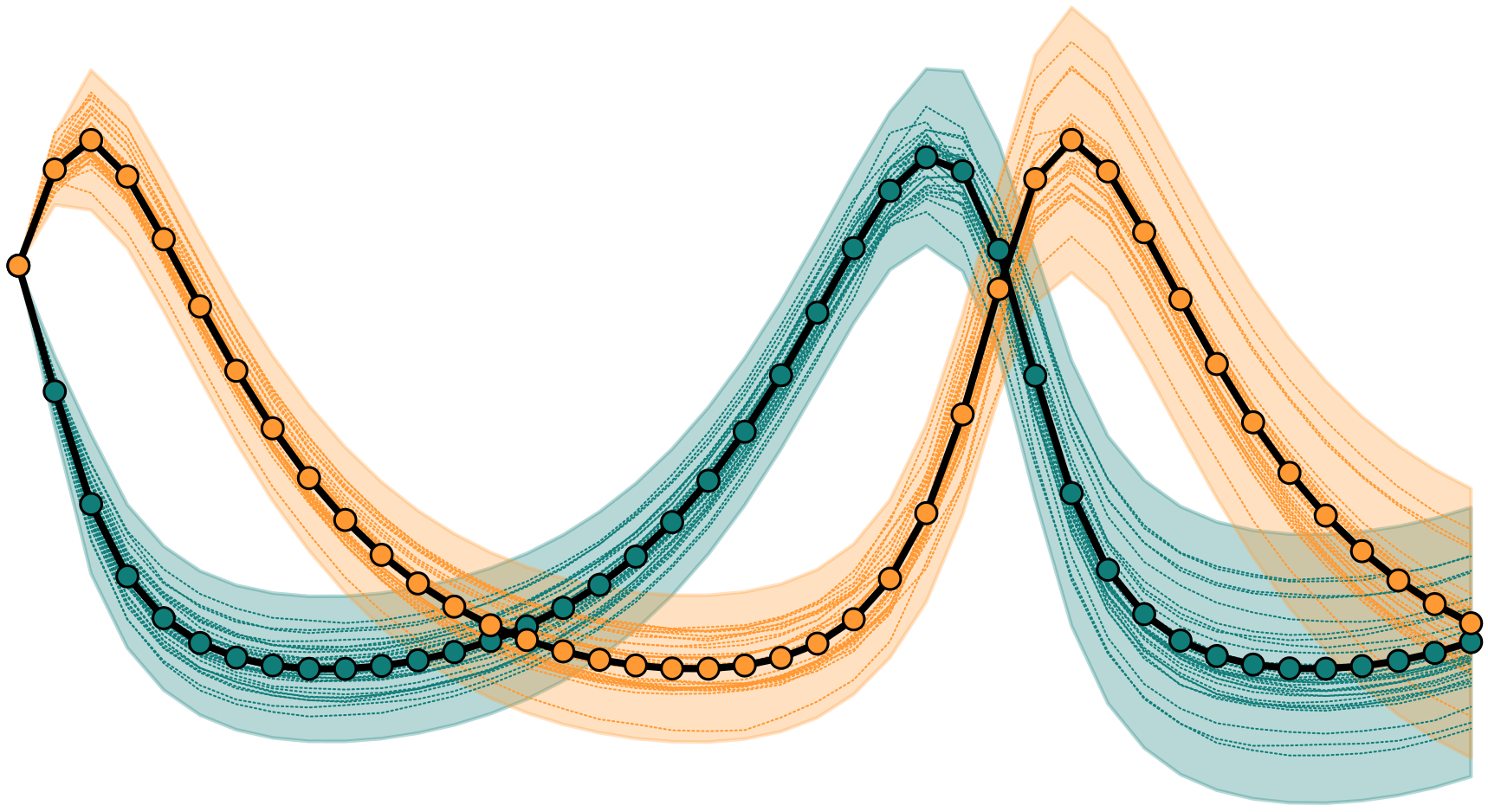}}
    \subcaption{Output belief of a PN method for ODEs.\label{fig:pnmethod-output}}
  \end{subfigure}
  \caption{\probnum{}'s probabilistic numerical methods compute beliefs over the solution of a numerical problem. The example code in (a) shows how to solve a linear system, while (b) illustrates the uncertainty arising from approximating the solution of a differential equation.}
  \label{fig:probnum-illustration}
\end{figure}

\section{Core Functionality}

\probnum{} implements probabilistic numerical methods (PNMs) for solving linear
systems,
differential equations, and integration problems (see \Cref{fig:probnum-illustration} for an example).
\Cref{tab:solvers} lists all currently
implemented solvers.

\begin{table}[h!]
  \caption{Probabilistic numerical methods implemented in
    \probnum{}.\label{tab:solvers}}
  \centering
  {\small
    \begin{tabular}{cccccc}
      \toprule
      \textbf{Area}                   & \textbf{Problem}                                       & \textbf{s.t.}                                      & \textbf{QoI}              & \textbf{Solver}      & \textbf{Ref.}                                        \\
      \midrule
      \multirow{2}{*}{Linear Algebra} & \multirow{2}{*}{\(Ax = b\)}                            & \(A \in \mathbb{R}^{n \times n}\)                  & \(x, A^{-1}\)             & Prob. linear solver  & \cite{Hennig2015,Wenger2020}                         \\
                                      &                                                        & \(A\) spd                                          & \(x\)                     & BayesCG              & \cite{Cockayne2019,Reid2021}                         \\
      \midrule
      \multirow{2}{*}{ODEs}           & \multirow{2}{*}{\(\dot{y}(t) = f(y(t), t)\)}           & \multirow{2}{*}{\(y(t_0) = y_0\)}                  & \multirow{2}{*}{\(y(t)\)} & ODE filter           & \cite{Kraemer2020,Bosch2021,Tronarp2019,Tronarp2021} \\
                                      &                                                        &                                                    &                           & Perturbed solver     & \cite{Abdulle2020}                                   \\
      \midrule
      \multirow{2}{*}{Integration}    & \multirow{2}{*}{\(F = \int_{\Omega} f(x) \, d\mu(x)\)} & \multirow{2}{*}{\(\Omega \subseteq \mathbb{R}^n\)} & \multirow{2}{*}{\(F\)}    & Bayesian Monte Carlo & \cite{Rasmussen2003}                                 \\
                                      &                                                        &                                                    &                           & Bayesian quadrature  & \cite{OHagan1991}                                    \\
      \bottomrule
    \end{tabular}}
\end{table}

\paragraph{Linear Algebra}
PNMs for linear algebra are focused on solving linear systems. \probnum{}
implements
iterative solvers, which either compute a belief over the (pseudo-)inverse of the matrix
\cite{Hennig2015,Wenger2020} or the solution directly \cite{Cockayne2019,Reid2021}. While these
perspectives differ conceptually, they are inherently
connected \cite{Bartels2019}.

\paragraph{Ordinary Differential Equations}
\probnum{}'s stable implementation \citep{Kraemer2020} of filtering-based
ODE solvers and their variations \citep{Tronarp2019,Tronarp2021} enables uncertainty calibration,
step-size adaptation, dense output, event handling and posterior sampling
\citep{Bosch2021,Kraemer2020}.
\probnum{} also provides perturbation-based ODE solvers
\cite{Abdulle2020}.

\paragraph{Numerical Integration}
The Bayesian quadrature implementation in \probnum{} comprises both
Bayesian Monte Carlo~\citep{Rasmussen2003} and Bayesian quadrature with fixed user-specified
points~\citep[e.g.][]{OHagan1991}. Currently, integrals can be estimated with respect to the
Lebesgue and Gaussian measure.

\section{Library Design}

\probnum{} mainly targets two groups of users. Those, that either want to
\begin{enumerate}[label=(\alph*)]
  \item \emph{use PNMs out-of-the-box} and explore their potential for their application of interest; or
  \item \emph{customize, develop and implement new PNMs} for specific problems.
\end{enumerate}
This focus on two different user groups is realized with a corresponding design pattern which
reflects \probnum{}'s core principles: \emph{usability},
\emph{extensibility}, and \emph{composability}.

\paragraph{Out-of-the-Box PN methods}
The widespread application of PN to scientific simulation problems necessitates that PN algorithms
can be called without knowing the algorithmic details of each solver.
Consequently, \probnum\ provides user-friendly interface methods for standard PN algorithms. These
take the
numerical problem to be solved -- and optionally known prior information -- as input and return a
belief over the quantities of interest. Interface methods often shadow the function signatures of
their direct \texttt{NumPy}
or \texttt{SciPy} equivalent, which makes them immediately \emph{usable}
inside more general simulation pipelines.

\begin{figure}
  \centering
  \adjustbox{max width=\textwidth, valign=c}{
    \input{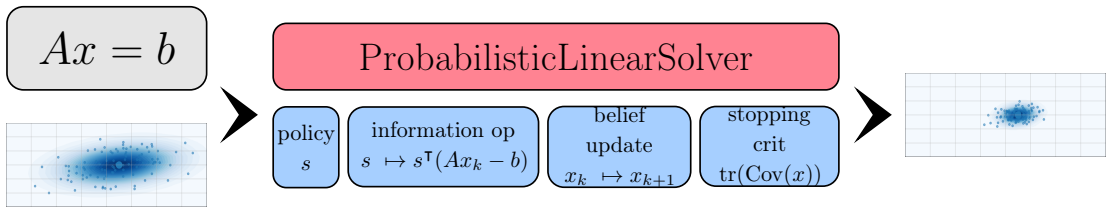}
  }
  \caption{Probabilistic numerical methods in \probnum{} are designed in a modular fashion. For example, a specialized linear solver (\colordot[gray,fill={rgb, 255:red, 255; green, 131; blue, 146 }]{4pt}) can be instantiated from (custom) modules (\colordot[gray,fill={rgb, 255:red, 166; green, 207; blue, 255 }]{4pt}). For a given problem \(Ax=b\) and prior belief, the solver then outputs a belief about the solution.}
  \label{fig:compositionality}
\end{figure}

\paragraph{Plug-and-Play for Custom Problems}
Often, specific problems require tailored numerical methods. Therefore, \probnum{}
allows the construction of PN methods from individual components.
In fact, \probnum{}'s interface methods merely call these lower-level,
customizable
implementations of PN methods. Each PNM is constructed from a set of components, for example,
policies, information operators, belief updates, hyperparameter optimization routines, or
perturbation strategies. An illustrative example is shown in \Cref{fig:compositionality}. This
compositional structure makes PN methods \emph{extensible}:
a user only needs to implement single components of PN algorithms in order to build a novel,
directly usable PN implementation. Finally, since all PN algorithm components act on random
variables or random processes, their
implementations are naturally \emph{composable} with each other.

\paragraph{Supporting Subpackages}
There are several \probnum{} subpackages such as basic data structures given by
random variables and processes, time- and memory-efficient matrix-free linear operators, as well as
Bayesian
filters and smoothers. These fundamentally enable and enhance the functionalities above, but may be
of independent interest.

\section{Project Development}

\probnum\ is an open-source, community-driven library hosted publicly on GitHub at
\begin{center}
  \vspace{-0.5em}
  \begin{tabular}{m{0.5cm} l}
    \includegraphics[width=0.05\textwidth]{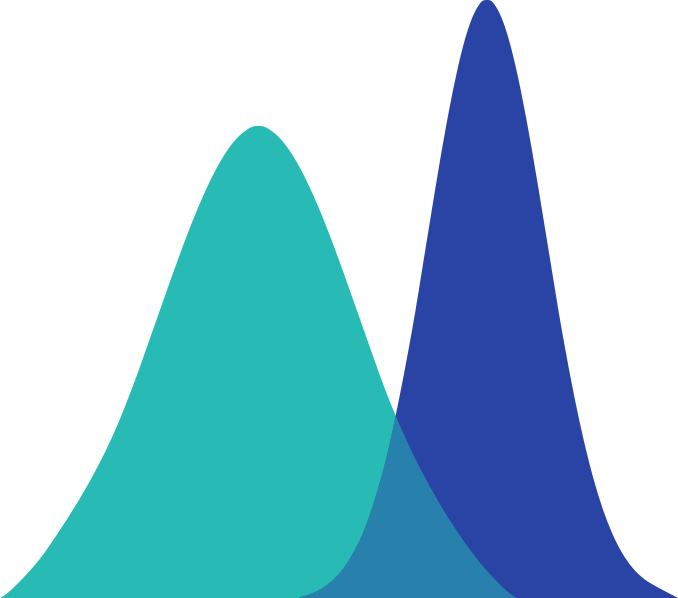} & \url{https://github.com/probabilistic-numerics/probnum}, \\
  \end{tabular}
  \vspace{-0.5em}
\end{center}
which allows for the tracking of issues, bugs and feature requests, as well as code contributions
via pull requests.
\probnum\ is distributed under the MIT license and available from the Python
Packaging Index (PyPI) for all recent Python versions via \texttt{pip install probnum}. All
contributors and community members are expected to adhere to the code of conduct.

\paragraph{Documentation} \probnum{}'s documentation can be found at
\url{www.probnum.org}. A range of tutorials showcasing PN methods, their applications, and
how
to use the library are also available.
A detailed development guide promotes contributions to the library.

\paragraph{Dependencies}
\probnum\ fundamentally only depends on \texttt{NumPy} and
\texttt{SciPy}, but includes the option to install additional packages to extend its
functionality (such as automatic differentiation frameworks). All dependencies are kept up-to-date
automatically via a bot.

\paragraph{Continuous Integration and Tests}
The project's documentation and tutorial notebooks are built for each commit to the main library
and every pull request.
Automatic linting preserves a high level of code quality and style.
\probnum{}'s extensive unit tests leverage
\texttt{pytest}, and the test coverage is updated automatically whenever new changes
occur. Finally, regular, automated benchmarks detect potential performance regressions.

\section{Related Concepts and Libraries}

\probnum{}'s solvers may be used as a drop-in replacement for
\emph{classic numerical solvers} (such as those offered by \texttt{SciPy} \citep{Virtanen2020}). While \probnum{} does not implement any classical
solvers, some PNMs fundamentally recover these without incurring
significant computational overhead
to compute uncertainty. The field of PN can be seen as a subset of the broad research area of
\emph{uncertainty quantification}, which ``is the
end-to-end study of the reliability of scientific inferences'' \cite{DepartmentEnergy2009}.
However,
existing software libraries for uncertainty quantification \citep{Debusschere2017,Olivier2020} do not
implement
any PN algorithms.
\probnum{} provides (approximate) inference algorithms for numerical tasks by
casting them as
statistical inference problems. It is therefore not a \emph{probabilistic programming} framework
\citep{Salvatier2016,Phan2019,Bingham2019}, which automate general probabilistic inference. Nevertheless,
\probnum{} may serve as a low-level building block inside probabilistic programs,
e.g.~to add computational uncertainty. Finally, there are special-purpose packages implementing
some PNMs, e.g. for Bayesian optimization \citep{Nogueira2014,Paleyes2019,Balandat2020} and quadrature
\citep{Paleyes2019,Choi2020}.

\section{Conclusion}

We hope that \probnum\ will increasingly be the home for algorithmic research in probabilistic
numerics and grow to further be a practical tool for applied users who want to leverage the
functionality of probabilistic numerical methods. For developers and researchers alike, \probnum\
offers the scaffold for efficient implementation of novel methods, as well as the repository of
existing methods for comparisons. For applied users, \probnum\ provides clean interfaces and a
``one-stop-shop'' for emerging new functionality.

\acks{We acknowledge contributions from Thomas Gläßle, Timothy Reid and others.\footnote{Please see \url{https://github.com/probabilistic-numerics/probnum/graphs/contributors} for a complete and continuously updated list of
    contributors.} The authors thank the participants of the Dagstuhl seminar
  ``Probabilistic Numerical Methods -- From Theory to Implementation'' for valuable discussions and
  feedback.  JW, NK, MP, JS, NB, MM, AG, and PH gratefully acknowledge financial support by the European Research Council through ERC StG
  Action 757275 / PANAMA; the DFG Cluster of Excellence ``Machine Learning - New Perspectives
  for Science'', EXC 2064/1, project number 390727645; the German Federal Ministry of Education and
  Research (BMBF) through the T\"ubingen AI Center (FKZ: 01IS18039A) as well as project ADIMEM (FKZ:
  01IS18052B); and funds from the Ministry of
  Science, Research and Arts of the State of Baden-W\"urttemberg.
  The authors thank the International Max Planck Research School for Intelligent Systems (IMPRS-IS)
  for supporting JW, NK, and NB.
  TK was supported by the Academy of Finland postdoctoral researcher grant 338567
  ``Scalable, adaptive and reliable probabilistic integration''.
}

{\small
  \bibliography{bibliography}

\begin{thebibliography}{29}
\providecommand{\natexlab}[1]{#1}
\providecommand{\url}[1]{\texttt{#1}}
\expandafter\ifx\csname urlstyle\endcsname\relax
  \providecommand{\doi}[1]{doi: #1}\else
  \providecommand{\doi}{doi: \begingroup \urlstyle{rm}\Url}\fi

\bibitem[Behrens and Dias(2015)]{Behrens2015}
J.~Behrens and F.~Dias.
\newblock New computational methods in tsunami science.
\newblock \emph{Philosophical Transactions of the Royal Society A:
  Mathematical, Physical and Engineering Sciences}, 373\penalty0 (2053), 2015.

\bibitem[Giles et~al.(2020)Giles, Kashdan, Salmanidou, Guillas, and
  Dias]{Giles2020}
Daniel Giles, Eugene Kashdan, Dimitra~M. Salmanidou, Serge Guillas, and
  Fr{\'{e}}d{\'{e}}ric Dias.
\newblock Performance analysis of {Volna-OP2} – massively parallel code for
  tsunami modelling.
\newblock \emph{Computers and Fluids}, 209, 2020.

\bibitem[Hennig et~al.(2015)Hennig, Osborne, and Girolami]{Hennig2015a}
Philipp Hennig, Mike~A. Osborne, and Mark Girolami.
\newblock Probabilistic numerics and uncertainty in computations.
\newblock \emph{Proceedings of the Royal Society of London A: Mathematical,
  Physical and Engineering Sciences}, 471\penalty0 (2179), 2015.

\bibitem[Cockayne et~al.(2019{\natexlab{a}})Cockayne, Oates, Sullivan, and
  Girolami]{Cockayne2019a}
Jon Cockayne, Chris Oates, Tim~J. Sullivan, and Mark Girolami.
\newblock Bayesian probabilistic numerical methods.
\newblock \emph{SIAM Review}, 61\penalty0 (4):\penalty0 756--789,
  2019{\natexlab{a}}.

\bibitem[Oates and Sullivan(2019)]{Oates2019}
Chris~J Oates and Tim~J Sullivan.
\newblock A modern retrospective on probabilistic numerics.
\newblock \emph{Statistics and Computing}, 29\penalty0 (6):\penalty0
  1335--1351, 2019.

\bibitem[Schmidt et~al.(2021)Schmidt, Kr{\"a}mer, and Hennig]{Schmidt2021}
Jonathan Schmidt, Nicholas Kr{\"a}mer, and Philipp Hennig.
\newblock A probabilistic state space model for joint inference from
  differential equations and data.
\newblock In \emph{Advances in Neural Information Processing Systems
  (NeurIPS)}, 2021.

\bibitem[Hennig(2015)]{Hennig2015}
Philipp Hennig.
\newblock Probabilistic interpretation of linear solvers.
\newblock \emph{SIAM Journal on Optimization}, 25\penalty0 (1):\penalty0
  234--260, 2015.

\bibitem[Wenger and Hennig(2020)]{Wenger2020}
Jonathan Wenger and Philipp Hennig.
\newblock Probabilistic linear solvers for machine learning.
\newblock In \emph{Advances in Neural Information Processing Systems
  (NeurIPS)}, 2020.

\bibitem[Cockayne et~al.(2019{\natexlab{b}})Cockayne, Oates, Ipsen, and
  Girolami]{Cockayne2019}
Jon Cockayne, Chris Oates, Ilse~C. Ipsen, and Mark Girolami.
\newblock A {B}ayesian conjugate gradient method.
\newblock \emph{Bayesian Analysis}, 14\penalty0 (3):\penalty0 937--1012,
  2019{\natexlab{b}}.

\bibitem[Reid et~al.(2021)Reid, Ipsen, Cockayne, and Oates]{Reid2021}
Tim~W. Reid, Ilse C.~F. Ipsen, Jon Cockayne, and Chris~J. Oates.
\newblock {BayesCG} as an uncertainty aware version of {CG}.
\newblock \emph{arXiv preprint}, 2021.
\newblock URL \url{http://arxiv.org/abs/2008.03225}.

\bibitem[Krämer and Hennig(2020)]{Kraemer2020}
Nicholas Krämer and Philipp Hennig.
\newblock Stable implementation of probabilistic {ODE} solvers.
\newblock \emph{arXiv preprint}, 2020.
\newblock URL \url{http://arxiv.org/abs/2012.10106}.

\bibitem[Bosch et~al.(2021)Bosch, Hennig, and Tronarp]{Bosch2021}
Nathanael Bosch, Philipp Hennig, and Filip Tronarp.
\newblock Calibrated adaptive probabilistic {ODE} solvers.
\newblock In \emph{Proceedings of The 24th International Conference on
  Artificial Intelligence and Statistics (AISTATS)}, 2021.

\bibitem[Tronarp et~al.(2019)Tronarp, Kersting, S{\"a}rkk{\"a}, and
  Hennig]{Tronarp2019}
Filip Tronarp, Hans Kersting, Simo S{\"a}rkk{\"a}, and Philipp Hennig.
\newblock Probabilistic solutions to ordinary differential equations as
  nonlinear {Bayesian} filtering: A new perspective.
\newblock \emph{Statistics and Computing}, 29\penalty0 (6):\penalty0
  1297--1315, 2019.

\bibitem[Tronarp et~al.(2021)Tronarp, S{\"a}rkk{\"a}, and Hennig]{Tronarp2021}
Filip Tronarp, Simo S{\"a}rkk{\"a}, and Philipp Hennig.
\newblock Bayesian {ODE} solvers: The maximum a posteriori estimate.
\newblock \emph{Statistics and Computing}, 31\penalty0 (3):\penalty0 1--18,
  2021.

\bibitem[Abdulle and Garegnani(2020)]{Abdulle2020}
Assyr Abdulle and Giacomo Garegnani.
\newblock Random time step probabilistic methods for uncertainty quantification
  in chaotic and geometric numerical integration.
\newblock \emph{Statistics and Computing}, 30:\penalty0 1--26, 2020.

\bibitem[Rasmussen and Ghahramani(2003)]{Rasmussen2003}
C.~E. Rasmussen and Z.~Ghahramani.
\newblock Bayesian {M}onte {C}arlo.
\newblock In \emph{Advances in Neural Information Processing Systems
  (NeurIPS)}, 2003.

\bibitem[O'Hagan(1991)]{OHagan1991}
A.~O'Hagan.
\newblock {B}ayes--{H}ermite quadrature.
\newblock \emph{Journal of Statistical Planning and Inference}, 29\penalty0
  (3):\penalty0 245--260, 1991.

\bibitem[Bartels et~al.(2019)Bartels, Cockayne, Ipsen, and Hennig]{Bartels2019}
Simon Bartels, Jon Cockayne, Ilse~C. Ipsen, and Philipp Hennig.
\newblock Probabilistic linear solvers: {A} unifying view.
\newblock \emph{Statistics and Computing}, 29\penalty0 (6):\penalty0
  1249--1263, 2019.

\bibitem[Virtanen et~al.(2020)Virtanen, Gommers, Oliphant, Haberland, Reddy,
  Cournapeau, Burovski, Peterson, Weckesser, Bright, et~al.]{Virtanen2020}
Pauli Virtanen, Ralf Gommers, Travis~E Oliphant, Matt Haberland, Tyler Reddy,
  David Cournapeau, Evgeni Burovski, Pearu Peterson, Warren Weckesser, Jonathan
  Bright, et~al.
\newblock {SciPy 1.0}: Fundamental algorithms for scientific computing in
  {Python}.
\newblock \emph{Nature Methods}, 17\penalty0 (3):\penalty0 261--272, 2020.

\bibitem[{U.S. Department of Energy}(2009)]{DepartmentEnergy2009}
{U.S. Department of Energy}.
\newblock Scientific grand challenges in national security: The role of
  computing at the extreme scale.
\newblock Technical report, 2009.

\bibitem[Debusschere et~al.(2017)Debusschere, Sargsyan, Safta, and
  Chowdhary]{Debusschere2017}
Bert Debusschere, Khachik Sargsyan, Cosmin Safta, and Kenny Chowdhary.
\newblock \emph{Uncertainty Quantification Toolkit (UQTk)}, pages 1807--1827.
\newblock Springer International Publishing, 2017.

\bibitem[Olivier et~al.(2020)Olivier, Giovanis, Aakash, Chauhan, Vandanapu, and
  Shields]{Olivier2020}
Audrey Olivier, Dimitrios Giovanis, B.~S. Aakash, Mohit Chauhan, Lohit
  Vandanapu, and Michael~D. Shields.
\newblock {UQpy}: a general purpose {P}ython package and development
  environment for uncertainty quantification.
\newblock \emph{Journal of Computational Science}, 47, 2020.

\bibitem[Salvatier et~al.(2016)Salvatier, Wiecki, and
  Fonnesbeck]{Salvatier2016}
John Salvatier, Thomas~V Wiecki, and Christopher Fonnesbeck.
\newblock Probabilistic programming in {Python} using {PyMC3}.
\newblock \emph{PeerJ Computer Science}, 2, 2016.

\bibitem[Phan et~al.(2019)Phan, Pradhan, and Jankowiak]{Phan2019}
Du~Phan, Neeraj Pradhan, and Martin Jankowiak.
\newblock Composable effects for flexible and accelerated probabilistic
  programming in {NumPyro}.
\newblock \emph{arXiv preprint}, 2019.
\newblock URL \url{http://arxiv.org/abs/1912.11554}.

\bibitem[Bingham et~al.(2019)Bingham, Chen, Jankowiak, Obermeyer, Pradhan,
  Karaletsos, Singh, Szerlip, Horsfall, and Goodman]{Bingham2019}
Eli Bingham, Jonathan~P Chen, Martin Jankowiak, Fritz Obermeyer, Neeraj
  Pradhan, Theofanis Karaletsos, Rohit Singh, Paul Szerlip, Paul Horsfall, and
  Noah~D Goodman.
\newblock {Pyro}: Deep universal probabilistic programming.
\newblock \emph{Journal of Machine Learning Research (JMLR)}, 20\penalty0
  (1):\penalty0 973--978, 2019.

\bibitem[Nogueira(2014)]{Nogueira2014}
Fernando Nogueira.
\newblock {Bayesian Optimization}: Open source constrained global optimization
  tool for {Python}, 2014.
\newblock URL \url{https://github.com/fmfn/BayesianOptimization}.

\bibitem[Paleyes et~al.(2019)Paleyes, Pullin, Mahsereci, Lawrence, and
  Gonz{\'a}lez]{Paleyes2019}
Andrei Paleyes, Mark Pullin, Maren Mahsereci, Neil Lawrence, and Javier
  Gonz{\'a}lez.
\newblock Emulation of physical processes with {Emukit}.
\newblock In \emph{Second Workshop on Machine Learning and the Physical
  Sciences (NeurIPS)}, 2019.

\bibitem[Balandat et~al.(2020)Balandat, Karrer, Jiang, Daulton, Letham, Wilson,
  and Bakshy]{Balandat2020}
Maximilian Balandat, Brian Karrer, Daniel~R. Jiang, Samuel Daulton, Benjamin
  Letham, Andrew~Gordon Wilson, and Eytan Bakshy.
\newblock {BoTorch}: A framework for efficient {M}onte-{C}arlo {B}ayesian
  optimization.
\newblock In \emph{Advances in Neural Information Processing Systems 33}, 2020.

\bibitem[Choi et~al.(2021)Choi, Hickernell, McCourt, and Sorokin]{Choi2020}
S.-C.~T. Choi, F.~J. Hickernell, M.~McCourt, and A.~Sorokin.
\newblock {QMCPy}: A quasi-{M}onte {C}arlo {P}ython library, 2021.
\newblock URL \url{http://arxiv.org/abs/2102.07833}.

\end{thebibliography}
}

\end{document}